\documentclass[12pt]{article}
\usepackage{amsmath}
\usepackage{amsfonts}
\usepackage{latexsym}
\usepackage{jstyle}
\usepackage{appendix}
\usepackage[utf8]{inputenc}

\newcommand{\tPsi}{{\tilde\Psi}}

\title{Symmetric formulation for higher spin correlators, quantum effective action and anomaly}

\author[]{Melik Karapetyan}
\author[]{and Ruben Manvelyan}

\affiliation[]{A. Alikhanyan National Laboratory (Yerevan Physics Institute) Alikhanian Br. Str. 2, 0036 Yerevan, Armenia}

\emailAdd{meliq.karapetyan@gmail.com}
\emailAdd{manvel@yerphi.am}

\abstract{We develop a constructive framework for the three-point higher spin conformal correlation function, originally introduced in our previous work, and apply it as the foundation for constructing the quantum effective action of the corresponding higher spin conformal gauge theory. Employing a symmetric refinement of the earlier construction, we analyze the principal singularity of the three-point function and the associated effective action. This leads to a general method for extracting the dominant singularity and investigating the anomalous local contributions to the effective action.}

\def \ollr{{\raise7pt\hbox{$\leftrightarrow  \! \! \! \! \! \!$}}}
\def \l{\langle}
\def \r{\rangle}

\def \hX{{\hat X}}
\def \hY{{\hat Y}}
\def \hZ{{\hat Z}}
\def \hI{{\hat I}}

\def \hT{{\hat \Theta}}

\def \hx{{\hat x}}

\def \ta{{\tilde a}}
\def \tb{{\tilde b}}
\def \tc{{\tilde c}}
\def \tC{{\tilde C}}

\def \tF{{\tilde F}}

\def \tG{{\tilde G}}

\def \A{{\cal A}}

\def \E{{\cal E}}

\def \I{{\cal I}}
\def \J{{\cal J}}

\def \L{{\cal L}}
\def \T{{\cal T}}

\def \hC{{\hat{C}}}

\begin{document}


\maketitle



\section{Introduction}

\indent

Higher-spin (HS) gauge theory encodes a large symmetry, both in the interacting case and even at the free level. This symmetry and the fact that HS  theories exist  in a flat and also in $AdS$ space make them suitable and interesting tools for verifying the $AdS/CFT$ duality \cite{Maldacena:1997re,Witten:1998qj}. 
On the bulk $AdS$ space for symmetric higher spin fields HS gauge symmetry helps to construct and classify cubic interaction vertices which is very important because we can say that for these weakly coupled HS gauge theories we have some information about interaction. Moreover, the last term in the $AdS$ cubic interaction is just the cubic interaction in flat Minkowski space and all the other terms with fewer derivatives can be interpreted  as curvature corrections. The relation of this weak bulk interaction to the strong coupled correlation functions for HS conserved  currents in conformal field theories on the boundary ($AdS/CFT$) is connected with the fact that conformal symmetry fixes the two- and three-point correlation functions up to a slight ambiguity represented by several constants.
We do not attempt to present an exhaustive list of references in this wide and still developing area, but rather refer the reader to a set of key references most relevant to our work. In the area of conformal correlators the shorter list is \cite{Polyakov:1970xd,Schreier:1971um,Migdal:1971fof,Ferrara:1973yt,Ruehl:1973nj,Ruehl:1973pr,
Koller:1974ut,Mack:1976pa,Osborn:1993cr,Osborn:1994rv,Erdmenger:1996yc,Park:1997bq,Osborn:1998qu,Park:1999pd,Anselmi:1999bb,Kuzenko:1999pi,
Park:1999cw,Giombi:2009wh,Giombi:2010vg,Giombi:2011rz,Costa:2011mg,Costa:2011dw,Maldacena:2011jn,Stanev:2012nq,Todorov:2012xx,Zhiboedov:2012bm,Alba:2013yda,
Costa:2014rya,Alba:2015upa,Kravchuk:2016qvl,Skvortsov:2018uru,Buchbinder:2022kmj,Buchbinder:2022mys,Buchbinder:2023coi,Karapetyan:2023zdu,Karapetyan:2025ick},  
with the second list for HS cubic interaction being somewhat longer \cite{Bengtsson:1983pd,Berends:1984wp,Berends:1984rq,Fradkin:1987ks,Fradkin:1986qy,Bengtsson:1986kh,Fradkin:1991iy,Metsaev:1991mt,
Metsaev:1991nb,Vasiliev:2001wa,Alkalaev:2002rq,Manvelyan:2004mb,Bekaert:2005jf,Metsaev:2005ar,Bekaert:2006us,Boulanger:2006gr,
Francia:2007qt,Fotopoulos:2007yq,Metsaev:2007rn,Fotopoulos:2008ka,Zinoviev:2008ck,Boulanger:2008tg,Manvelyan:2009tf,
Manvelyan:2009vy,Bekaert:2009ud,Manvelyan:2010wp,Manvelyan:2010jr,Sagnotti:2010at,Zinoviev:2010cr,Fotopoulos:2010ay,
Manvelyan:2010je,Polyakov:2010sk,Ruehl:2011tk,Vasiliev:2011knf,Joung:2011ww,Dempster:2012vw,Joung:2012rv,Buchbinder:2012iz,
Henneaux:2012wg,Joung:2012fv,Manvelyan:2012ww,Joung:2012hz,Boulanger:2012dx,Henneaux:2013gba,Joung:2013nma,Conde:2016izb,
Bengtsson:2016hss,Francia:2016weg,Taronna:2017wbx,Roiban:2017iqg,Sleight:2017pcz,Sleight:2017cax,Karapetyan:2019psg,Joung:2019wbl,
Fredenhagen:2019lsz,Khabarov:2020bgr,Karapetyan:2021wdc}.

Thus, in this case $AdS/CFT$ works as a one-to-one correspondence between cubic vertices in $d+1$-dimensional Minkowski space\cite{Metsaev:2005ar,Manvelyan:2010jr} and conformal correlators in $d$ dimensions. This was first investigated in \cite{Costa:2011mg,Costa:2011dw,Zhiboedov:2012bm}. In our two previous works \cite{Karapetyan:2023zdu},\cite{Karapetyan:2025ick} we reconsidered the problem in the Osborn-Petkou general formulation \cite{Osborn:1993cr}. Moreover in our second paper \cite{Karapetyan:2025ick} we proposed a constructive approach for the direct construction of the higher-spin conserved correlation functions using all possible products of the terms of correlation functions for spin-one and spin-two.
Actually, we formulated it in the general case and proved for spins $s=3$ and $4$ that the knowledge about spin-one and spin-two is enough for constructing the higher spin correlation function satisfying all the symmetry conditions for the structural tensor in Osborn-Petkou formulation \cite{Osborn:1993cr},\cite{Karapetyan:2023zdu}.

In this paper we develop our constructive approach in the direction of constructing of the  quantum cubic effective action. We investigate main singularity of three point function and effective action. We then we  develop  general method of extraction of the main singularity and investigate singular local part of effective action. Next we consider general mechanism of production of possible trace and gauge anomaly in specific dimensional regularization scheme formulated 
in \cite{Ruehl:2008bm} for another reason. To do that we consider first symmetric formulation of our constructive approach to get shortest way to singularity extraction and classification. This leads us to the result that our scheme breaks both tracelessness  and conservation condition (gauge invariance of effective action) but the source of these violations is the same: it is violation of tracelessness at the quantum level for local singular part of effective action. Finally we obtain full general description of anomaly emergence mechanism. This consideration works in dimension $d=4$ because it is connected with conservation condition of three point function and correspondingly with gauge variation of cubic quantum effective action. As a result we obtain that the anomaly is of second order in the linearized HS gauge field and contains $2s$ derivatives. It means that anomaly can be expressed as a combination of squares of the generalized Weyl and Ricci tensors and scalar. But this is true in dimensions $d=4$. For the investigation of the anomaly in dimensions higher than $4$
we need higher-point correlation functions, the structure of which is not fixed by conformal invariance.  
To finish this short introduction we briefly review the content of our paper:
 
In Section \ref{sec:two} we present a brief technical review of the formalism for working with higher spins and  for constructing the two and three point correlation function in the case of  coinciding (equal) spins. At the same time we present the general formula for quantum effective action and connection of the latter with correlation functions. 
In Section \ref{sec:three} we develop symmetric approach for formulation of correlators and correspondingly for symmetric formulation of the cubic effective action.
We then formulate gauge invariance of effective action and connection with conservation condition for correlation function. The example of spin-two is presented in details in Appendix \ref{app:A}. 
In Section \ref{sec:four} we briefly present the main singularity-extraction procedure proposed in \cite{Ruehl:2008bm}. All technical details and derivations from this section are included in detail in Appendix \ref{app:B}.
Preparing the main ingredients we turn in Section \ref{sec:five} to splitting our cubic effective action into singular and regular
parts. After that we understand that anomaly can be obtained if symmetry of \emph{local singular} part violated by \emph{local finite} term and this is the source of the quantum anomaly. Finally we obtain that at the quantum level we will violate both conditions for our conserved current: tracelessness and conservation. But the violation of the conservation condition and the nonzero quantum trace are connected by a gradient. As a result, we can restore gauge invariance by standard shifting of currents keeping only trace anomaly at the quantum level of classical conformal field theory. 
We finish the paper with a short conclusion and outlook.

\section{Effective action and conformal correlation functions}\label{sec:two}
\renewcommand{\theequation}{\arabic{section}.\arabic{equation}}\setcounter{equation}{0}

\indent

We deal with HS fields in the same way as in our previous articles, namely by using additional auxiliary vector variables $a_{\mu}, b_{\mu},\dots$  to operate an arbitrary number of symmetrized indices. In this approach instead of symmetric tensors such as $h^{(s)}_{\mu_1\mu_2...\mu_s}(x)$ we work with the homogeneous polynomials in a vector $a^{\mu}$ of degree $s$ at the space-time point $x$:
\begin{equation}\label{2.1}
h^{(s)}(a;x) = h^{(s)}_{\mu_1\mu_2...\mu_s}(x)a^{\mu_{1}}a^{\mu_{2}}\dots a^{\mu_{s}} .
\end{equation}
Then the symmetrized gradient, divergence, and trace operations are given as usual:
\begin{eqnarray}
&&Grad:h^{(s)}(x;a)\Rightarrow (Grad\, h)^{(s+1)}(x;a) = (a\nabla)h^{(s)}(x;a)\,, \label{2.2}\\
&&Div:h^{(s)}(x;a)\Rightarrow (Div\, h)^{(s-1)}(x;a) = \frac{1}{s}(\nabla\partial_{a})h^{(s)}(x;a)\,,\label{2.3}\\
&&Tr:h^{(s)}(x;a)\Rightarrow (Tr\, h)^{(s-2)}(x;a) = \frac{1}{s(s-1)}\Box_{a}h^{(s)}(x;a)\,.\label{2.4}
\end{eqnarray}
In order not to confuse the derivatives in the $a$ and $x$ spaces, we define the notation $\nabla_{\mu}$ for space-time derivatives $\frac{\partial}{\partial x^{\mu}}$ preserving the notation $\partial_{a}$ for derivative in the auxiliary space.
Other useful notations are ``star" products  $*_a, *_b,\dots$ for a full contraction of $s$ symmetric indices:
\begin{eqnarray}
  *^{(s)}_{a}&=&\frac{1}{(s!)^{2}} \prod^{s}_{i=1}\overleftarrow{\partial}^{\mu_{i}}_{a}\overrightarrow{\partial}_{\mu_{i}}^{a} . \label{2.5}
\end{eqnarray}
These operators shorten drastically tensorial notation of the correlation functions  of higher spin currents and corresponding HS gauge fields in construction of the quantum effective action. 

Now we can define the general quantum effective action for HS fields through the nonlocal  correlation functions:
\begin{align}
  W^{eff}(h^{(s)})&=\sum_{k}\int_{x_{1}}\dots\int_{x_{k}}\l \prod^{k}_{i=1}\J^{(s)}(a_{i};x_{i})\r \prod^{k}_{i=1}\left\{*^{(s)}_{a_{i}}h^{(s)}(a_{i};x_{i})dx_{i}\right\}\label{2.6}
\end{align}
where
\begin{align}
  \l \J^{(s)}(a_{1};x_{1})\J^{(s)}(a_{2};x_{2})\dots \J^{(s)}(a_{k};x_{k})\r \label{2.7}
\end{align}
is $k$-point conformal correlation function of spin $s$ traceless conserved currents.
In reality we have a crucial restriction coming from the well-known fact that conformal symmetry fixes the form of the two- and three-point correlation functions. Therefore we can consider our effective action up to cubic level on the gauge fields. Fortunately, this cubic level of consideration is enough for investigation of HS gauge and HS conformal anomalies in dimension $d=4$. 
So, the action we are going to study is built from two- and three-point correlators. These correlation functions for higher spin fields have been investigated in various ways in the literature. In particular, in our recent works we classified and analyzed the three-point correlators for conserved currents of arbitrary spin and proposed a constructive approach for building the correlators from the solutions for spin-one and spin-two. An important aspect of this construction was the use of the Osborn-Petkou approach, where the structural tensor of the correlators is a function of a single variable, which significantly simplifies the construction procedure, including within the constructive framework, and also facilitates solving the conservation condition. However, in this work we are interested in studying the singular behavior of correlators and the effective action, and then the emergence of an anomaly in the dimensional regularization scheme, which, as we will see later, is much more convenient to consider in the symmetric formulation, where the correlators are explicitly symmetric with respect to all points. Consequently, the procedure of variation with respect to the external field and the verification of symmetry become greatly simplified. 
Here we review the construction of two and three point correlation functions for HS case. The two-point function is initially symmetric and is defined up to one constant:
\begin{align}
  \big<\J^{(s)}(a;x_{1}) \J^{(s)}(b;x_{2}) \big>&=\frac{C_{\J}}{(x_{12}^{2})^{\Delta_{(s)}}}\I^{(s)}(a,b;x_{12})\label{2.8}\\
  x^{\mu}_{12}&=x^{\mu}_{1}-x^{\mu}_{2}\label{2.9}
\end{align}

The $\I^{(s)}(a;x_{i})$ in (\ref{2.8}) is just $s$-th power of the inversion matrix  
\begin{align}
 I(a,b;x)=(ab)-2(a\hx)(b\hx),  \quad \hx_{\mu}=\frac{x_{\mu}}{\sqrt{x^{2}}} \label{2.10}
\end{align}
projected to the space of traceless tensors:
\begin{align}
  &\I^{(s)}(a,b;x)=\big(I(a,c;x)\big)^{s} *^{s}_{c}\E^{(s)}(c,b)=\E^{(s)}(a,c)*^{s}_{c}(I(c,b;x))^{s}\label{2.11}\\
  &\Box_{a,b} \I^{(s)}(a,b;x)=0 ,\label{2.12}
\end{align}
 here $\E^{(s)}(a,c)$ is the projector onto the space of traceless rank-$s$ tensors:
\begin{align}
  &T^{(s)}_{traceless}(a)=\E^{(s)}(a,b)*^{(s)}_{b}T^{(s)}(b)\label{2.13}\\
 &\Box_{a}\E^{(s)}(a,b)=  \Box_{b}\E^{(s)}(a,b)=0\label{2.14}
\end{align}
 The exact form of this projector is worth presenting here:
\begin{align}
  \E^{(s)}(a,b)=\sum^{s/2}_{p=0}\lambda_{p}(ab)^{s-2p}(a^{2}b^{2})^{p} , \quad \lambda_{0}=1\label{2.15}
\end{align}
where the coefficients $\{\lambda_{p}\}^{s/2}_{p=0}$ are  determined by the tracelessness condition (\ref{2.14}) together with normalization condition for $\lambda_{0}$ in (\ref{2.15})\footnote{Here we use notations $[a]_{n}$ for falling factorials (Pochhammer symbols):
\begin{align*}
[a]_{n}=\frac{a!}{(a-n)!}=\frac{\Gamma(a+1)}{\Gamma(a-n+1)}
\end{align*}
 see \cite{Karapetyan:2023zdu} for details.}
\begin{align}
\lambda_{p}=\frac{(-1)^{p}[s]_{2p}}{2^{2p}p![d/2+s-2]_{p}}\label{2.16}
\end{align}
Three-point correlation functions for spin-one and spin-two conserved currents are formulated in detail in \cite{Osborn:1993cr}. The short review can be found in \cite{Karapetyan:2023zdu}.
The result of this formulation for the general three-point function of three equal traceless
higher-spin-s currents was presented in our previous papers are:
\begin{align}
&\l \J^{(s)}(a;x_{1}) \,\J^{(s)}(b;x_2) \, \J^{(s)}(c;x_3) \r = \nonumber\\
&=\frac{1}{ x_{12}^{\Delta_{(s)}} x_{23}^{\Delta_{(s)}} x_{31}^{\Delta_{(s)}}}\I^{(s)}(a,a';x_{13}) \I^{(s)}(b,b';x_{23})*^{(s)}_{a'}*^{(s)}_{b'}t^{(s)}(a',b';c;\hX_{12}), \label{2.17}
\end{align}
where hatted coordinates
\begin{align}
\hX_{\mu}=\frac{X_{\mu}}{\sqrt{X^{2}}}\,\,\label{2.18}
\end{align}
are unit vectors and
\begin{align}
\hX_{12\mu} = -\hX_{21\mu} = \sqrt{\frac{x_{13}^{\, 2} x_{23}^{\, 2}}{x_{12}^{\, 2}}}\left[\frac{x_{13\mu}}{x_{13}^{\, 2}} -
\frac{x_{23\mu}}{x_{23}^{\, 2}}\right].\label{2.19}
\end{align}
The scaling dimension $\Delta_{(s)}$ of the spin-$s$ currents appearing in the two- and three-point functions is fixed by the conservation condition::
\begin{align}\label{2.20}
 \Delta_{(s)}=d+s-2 ,
\end{align}
which is correct for the dual massless higher-spin gauge field.

The principal object of this formulation is the structural tensor $t^{(s)}(a,b;c;\hX_{12})$.
The tracelessness of the $t^{(s)}(a,b;c;\hX_{12})$ in all three sets of symmetrized indices can be used to define a  ``\emph{kernel}'' object $\tilde{t}^{(s)}(a,b;c;\hX)$ enveloped by three traceless projectors
\begin{align}\label{2.21}
t^{(s)}(\ta,\tb;\tc;\hX)=\E^{(s)}(\ta,a)*_{a}\E^{(s)}(\tb,b)*_{b}\tilde{t}^{(s)}(a,b;c;\hX)*_{c}\E^{(s)}(c,\tc).
\end{align}
In that case the standard symmetry properties of $t^{(s)}(a,b;c;\hX_{12})$ and the conservation condition defined in \cite{Osborn:1993cr} can be formulated for the ``kernel'' $\tilde{t}^{(s)}(a,b;c;\hX)$. The useful point here that structural tensor and its ``kernel'' are function of one variable only. So this allows us to develop an easier procedure for the construction and classification of the s+1 solutions of the conservation condition for the general spin $s$ case.  Even for the correlators with different spins \cite{Karapetyan:2023zdu} we can perform this procedure and arrive at the $s_{min}+1$ solutions of conservation condition. Moreover in \cite{Karapetyan:2025ick} we developed new constructive approach for construction of the structural tensor for general spin from the spin-one and spin-two principal tensor objects. 

The key point of the constructive approach is that, starting from two solutions of the conservation condition for the structural tensor of spin-one:
\begin{align}
  \tilde{t}^{(s=1)}_{1}(a,b;c;\hX) &= G(a,b;c;\hX), \label{2.22}\\
 \tilde{t}^{(s=1)}_{2}(a,b;c;\hX) &= \Psi(a,b,c;\hX),\label{2.23}
\end{align}
where
\begin{align}
  G(a,b;c;\hX)& =(a\hX)(bc)+(b\hX)(ac)-(c\hX)(ab),\label{2.24}\\
   \Psi(a,b,c;\hX)& = (a\hX)(b\hX)(c\hX).\label{2.25}
\end{align}
We can construct only one solution of the conservation condition for spin $s=2$~\cite{Karapetyan:2025ick}.
\begin{align}
 & \tilde{t}^{(s=2)}_{1}(a,b;c;\hX) =G^{(2)}(a,b;c;\hX)+(d+2)G(a,b;c;\hX)\Psi(a,b,c;\hX)\label{2.26}
\end{align}
The remaining two solutions
 \begin{align}
 \tilde{t}^{(s=2)}_{2}(a,b;c;\hX)& = {\cal F }_{1}(a,b;c;\hX)\nonumber\\
 &-\frac{d(d+2)-8}{2}\left(G(a,b;c;\hX)\Psi(a,b,c;\hX)-\Psi^{2}(a,b,c;\hX)\right),\label{2.27}\\
 \tilde{t}^{(s=2)}_{3}(a,b;c;\hX)&={\cal F }_{2}(a,b;c;\hX)-{\cal F }_{1}(a,b;c;\hX)-d^{2}G(a,b;c;\hX)\Psi(a,b,c;\hX)\label{2.28}
\end{align}
require the introduction of two more fundamentally independent objects forming the spin-two conserved structural tensors \cite{Karapetyan:2025ick}. 
 \begin{align}
  &{\cal F }_{1}(a,b;c;\hX)= (ab)I(a,c;\hX)I(b,c;\hX),\label{2.29}\\
  &{\cal F }_{2}(a,b;c;\hX)= (ab)(ac)(bc)+(bc)I(a,b;\hX)I(a,c;\hX)+(ac)I(a,b;\hX)I(b,c;\hX). \label{2.30}
\end{align}
These correspond to curvature corrections in the dual cubic interaction on the bulk\footnote{We refer to our previous papers \cite{Karapetyan:2023zdu}, \cite{Karapetyan:2025ick} for correct detailed formulation of the  conservation condition and the correct symmetry properties of all objects.} \cite{Karapetyan:2023zdu}, \cite{Karapetyan:2025ick}. 
The most remarkable aspect of the constructive approach is that these four quantities:
\begin{align}
 G(a,b;c;\hX);\,\,\Psi(a,b;c;\hX)\;\,\,;{\cal F }_{1}(a,b;c;\hX);\,\, {\cal F }_{2}(a,b;c;\hX), \label{2.31}
\end{align}
are sufficient to form the structural tensor for the three-point equal-spin  correlation function for any spin:
\begin{align}
&\tilde{t}^{(s)}(a,b;c;\hX)\nonumber\\
&=\sum^{[s/2]}_{k=0}\sum^{s-2k}_{n=0}\sum^{k}_{m=0}A_{knm} G(a,b;c;\hX)^{s-2k-n}\Psi(a,b;c;\hX)^{n}{\cal F }_{1}(a,b;c;\hX)^{k-m}{\cal F }_{2}(a,b;c;\hX)^{m},\label{2.32}
\end{align}
with  the correct number of independent coefficients
\begin{align}\label{2.33}
  \# (A_{knm})=\sum^{[s/2]}_{k=0}(s-2k+1)(k+1),
\end{align}
which is in agreement with our general results in \cite{Karapetyan:2023zdu}.
Then the conservation condition
\begin{align}
 & (\nabla_{x_{1}}\partial_{a})\langle\J^{(s)}(a;x_{1}) \,\J^{(s)}(b;x_2) \, \J^{(s)}(c;x_3) \r =0\label{2.34}
\end{align}  
leads to the $s+1$ surviving combinations for structural tensor. We proved this in detail for spins three and four in \cite{Karapetyan:2025ick}.
\section{Symmetric formulation of correlators and effective action}\label{sec:three}
Thus, we draw two main conclusions from the previous section: first, that we must restrict ourselves to the part of the effective action determined by the two- and three-point correlation functions, and second, that it is necessary to develop a procedure for transformation to a symmetric formulation of the
three-point function and structural tensor in order to study the behavior of the effective action.
First of all, we introduce some new notation\footnote{These notations are similar to \cite{Erdmenger:1996yc}}
\begin{align}
&x^{\mu}=x^{\mu}_{23}=x^{\mu}_{2}-x^{\mu}_{3},\label{3.1}\\
&y^{\mu}=x^{\mu}_{31}=x^{\mu}_{3}-x^{\mu}_{1},\label{3.2}\\
&z^{\mu}=x^{\mu}_{12}=x^{\mu}_{1}-x^{\mu}_{2};\label{3.3}\\
&x^{\mu}+y^{\mu}+z^{\mu}=0\label{3.4}
\end{align}
The last relation, of course, means translational invariance in the notation $z,y,x$.
Then instead of the standard notation of inverted differences \cite{Osborn:1993cr} we define like in \cite{Erdmenger:1996yc} the capital $X,Y,Z$
\begin{align}
&X^{\mu}=-\frac{z^{\mu}}{z^{2}}-\frac{y^{\mu}}{y^{2}}=X^{\mu}_{23}; \label{3.5}\\
&Y^{\mu}=-\frac{x^{\mu}}{x^{2}}-\frac{z^{\mu}}{z^{2}}=X^{\mu}_{31};\label{3.6}\\ 
&Z^{\mu}=-\frac{y^{\mu}}{y^{2}}-\frac{x^{\mu}}{x^{2}}=X^{\mu}_{12}\label{3.7}
\end{align}
then we can calculate
\begin{align}
  X^{2}=\frac{x^{2}}{y^{2}z^{2}},\quad
  Y^{2}=\frac{y^{2}}{x^{2}z^{2}}, \quad
  Z^{2}=\frac{z^{2}}{x^{2}y^{2}}  \label{3.8}
\end{align}
and then define the corresponding unit vectors $\hX^{\mu}, \hY^{\mu},\hZ^{\mu}$.
We can then derive three sets of relations that are very important for symmetrization.
The first set we call the  ``cyclic transition rules", taking us from the small $x,y,z$ to the capital $X,Y,Z$  in the language of inversion matrices:
\begin{align}
  I^{\lambda}_{\mu}(x)I_{\lambda \nu}(z)=I^{\lambda}_{\mu}(y)I_{\lambda \nu}(\hX),
  \quad I^{\lambda}_{\mu}(x)I_{\lambda \nu}(y)=I^{\lambda}_{\mu}(z)I_{\lambda \nu}(\hX),\label{3.9}\\
  I^{\lambda}_{\mu}(y)I_{\lambda \nu}(x)=I^{\lambda}_{\mu}(z)I_{\lambda \nu}(\hY),
  \quad I^{\lambda}_{\mu}(y)I_{\lambda \nu}(z)=I^{\lambda}_{\mu}(x)I_{\lambda \nu}(\hY),\label{3.10}\\
  I^{\lambda}_{\mu}(z)I_{\lambda \nu}(y)=I^{\lambda}_{\mu}(x)I_{\lambda \nu}(\hZ),
  \quad I^{\lambda}_{\mu}(z)I_{\lambda \nu}(x)=I^{\lambda}_{\mu}(y)I_{\lambda \nu}(\hZ)\label{3.11}
\end{align}
The second set consists of the ``cyclic rotation rules":
\begin{align}
  I^{\lambda}_{\mu}(x)\hZ_{\lambda}=-\hY_{\mu},\quad I^{\lambda}_{\mu}(y)\hZ_{\lambda}=-\hX_{\mu},\label{3.12}\\
  I^{\lambda}_{\mu}(x)\hY_{\lambda}=-\hZ_{\mu},\quad I^{\lambda}_{\mu}(z)\hY_{\lambda}=-\hX_{\mu},\label{3.13}\\
  I^{\lambda}_{\mu}(y)\hX_{\lambda}=-\hZ_{\mu},\quad I^{\lambda}_{\mu}(z)\hX_{\lambda}=-\hY_{\mu},\label{3.14}
\end{align}
and the third set consists of the ``cyclic shifting rules", obtained from the first two by trivial combinations:
\begin{align}
  I^{\lambda}_{\mu}(x)I_{\lambda \nu}(z)=I_{\mu\nu}(y)+2\hZ_{\mu}\hX_{\nu},
  \quad I^{\lambda}_{\mu}(z)I_{\lambda \nu}(x)=I_{\mu\nu}(y)+2\hX_{\mu}\hZ_{\nu},\label{3.15}\\
  I^{\lambda}_{\mu}(y)I_{\lambda \nu}(x)=I_{\mu\nu}(z)+2\hX_{\mu}\hY_{\nu},
  \quad I^{\lambda}_{\mu}(x)I_{\lambda \nu}(y)=I_{\mu\nu}(z)+2\hY_{\mu}\hX_{\nu},\label{3.16}\\
  I^{\lambda}_{\mu}(z)I_{\lambda \nu}(y)=I_{\mu\nu}(x)+2\hY_{\mu}\hZ_{\nu},
  \quad I^{\lambda}_{\mu}(y)I_{\lambda \nu}(z)=I_{\mu\nu}(x)+2\hZ_{\mu}\hY_{\nu},\label{3.17}
\end{align}

Now we can use these rules for symmetrization of all primary objects  which form the structural tensor.
We can show that:
\begin{align}
&\tG(a,b,c;\hZ,\hY,\hX)=I(b,b';x)I(a,a';y)*^{(1)}_{a'}*^{(1)}_{b'}G(a',b',c;\hZ)\nonumber\\
&=-(a\hX)I(b,c;x)-(b\hY)I(a,c;y)-(c\hZ)I(a,b;z)-2(a\hX)(b\hY)(c\hZ)\label{3.18}\\
&\tPsi(a,b,c;\hZ,\hY,\hX)=I(b,b';x)I(a,a';y)*^{(1)}_{a'}*^{(1)}_{b'}\Psi(a,b,c;\hZ)=(a\hX)(b\hY)(c\hZ)\label{3.19}\\
&{\cal{\tF}}_{1}(a,b,c;\hZ,\hY,\hX)=I^{2}(b,b';x)I^{2}(a,a';y)*^{(2)}_{a'}*^{(2)}_{b'}{\cal F}_{1}(a,b;c;\hZ)\nonumber\\
&=[I(a,b;z)+2(a\hX)(b\hY)][I(a,c;y)+2(a\hX)(c\hZ)][I(b,c;x)+2(b\hY)(c\hZ)]\label{3.20}\\
&{\cal{\tF}}_{2}(a,b,c;\hZ,\hY,\hX)=I^{2}(b,b';x)I^{2}(a,a';y)*^{(2)}_{a'}*^{(2)}_{b'}{\cal F}_{2}(a,b;c;\hZ)\nonumber\\
&=[I(a,b;z)+2(a\hX)(b\hY)]I(a,c;y)I(b,c;x)+I(a,b;z)[I(a,c;y)+2(a\hX)(c\hZ)]I(b,c;x)\nonumber\\
&+I(a,b;z)I(a,c;y)[I(b,c;x)+2(b\hY)(c\hZ)]\label{3.21}
\end{align}
So we see that we can absorb all the inversion matrices of the three-point function
\begin{align}
  \frac{1}{ z^{\Delta_{(s)}} x^{\Delta_{(s)}} y^{\Delta_{(s)}}} I^{s}(a,a';y) I^{s}(b,b';x)*^{(s)}_{a'}*^{(s)}_{b'}\tilde{t}^{(s)}(a',b';c;\hZ) \label{3.22}
\end{align}
in structural tensor $\tilde{t}^{(s)}(a',b';c;\hZ)$ constructed from spin-one and spin-two principal terms in our approach.

Finally we obtain symmetric expression:
\begin{align}
  \frac{1}{ z^{\Delta_{(s)}} x^{\Delta_{(s)}} y^{\Delta_{(s)}}}\tilde{t}^{(s)}(a,b,c;\hZ,\hY,\hX)\label{3.23}
\end{align}

Where the symmetric structural tensor $\tilde{t}^{(s)}(a,b,c;\hZ,\hY,\hX)$ is obtained from the usual one $\tilde{t}^{(s)}(a',b';c;\hZ)$ by the following rules:
\begin{align}
 &G(a',b',c;\hZ)\Rightarrow \tG(a,b,c;\hZ,\hY,\hX) ;\quad\quad \Psi(a,b,c;\hZ) \Rightarrow \tPsi(a,b,c;\hZ,\hY,\hX)\label{3.24}\\
 &{\cal F}_{1}(a,b;c;\hZ)\Rightarrow {\cal \tF}_{1}(a,b,c;\hZ,\hY,\hX); \quad\quad {\cal F}_{2}(a,b;c;\hZ)\Rightarrow {\cal \tF}_{2}(a,b,c;\hZ,\hY,\hX)\label{3.25}
\end{align}
In Appendix \ref{app:A} we present the detailed and optimized construction of the three-point correlation function in the symmetric formulation for spin $2$. From this consideration, the structure, the conservation condition and the tracelessness of the singular part of the effective action become more obvious.
Summarizing this consideration we can write our effective action up to cubic order in the following form
\begin{align}
  &W^{eff}(h^{(s)})=C_{\J}\int\int dx_{1}dx_{2}\frac{I^{s}(a,b;z)}{(z^{2})^{\Delta_{(s)}}}*^{s}_{a}*^{s}_{b}h^{T}(a;x_{1})h^{T}(b;x_{2})\nonumber\\
  &+\int\int\int dx_{1}dx_{2}dx_{3}\frac{\tilde{t}^{(s)}(a,b,c;\hZ,\hY,\hX)}{ z^{\Delta_{(s)}} x^{\Delta_{(s)}} y^{\Delta_{(s)}}}*^{s}_{a}*^{s}_{b}*^{s}_{c}h^{T}(a;x_{1})h^{T}(b;x_{2})h^{T}(c;x_{3}) \label{3.26}
\end{align}
where we write down only the ``kernels" of the correlation functions and hide all the projection operators in the definition of traceless external fields:
\begin{align}
  h^{T}(a_{i};x_{i})=\E^{(s)}(a_{i},\ta_{i})*^{s}_{\ta_{i}}h(\ta_{i};x_{i})\label{3.27}
\end{align}
Next important formula is connected with gauge variation of the effective action. 
After a rather long but straightforward calculation we can derive the following relation:
\begin{align}
  (\ta\nabla)*^{s}_{\ta}\E^{(s)}(\ta, a)=\E^{(s-1)}(\ta, a)*^{s-1}_{a}\left[(\nabla\partial_{a})-\frac{1}{d+2s-4}(a\nabla)\Box_{a}\right] \label{3.28}
\end{align}
Hence, the gauge variation of the gauge fields:
\begin{align}
 \delta h^{(s)}(\ta;x)=(\ta\nabla)\alpha^{(s-1)}(\ta;x)\label{3.29}
\end{align}
leads to the following conservation conditions for ``kernels" of correlators\footnote{The presence of two remaining projectors in $b$ and $c$ space ensures that the conservation condition (\ref{3.31}) for the kernel is satisfied up to $b^{2}, c^{2}$ terms. This is also true for (\ref{3.30}), but only in $b$ space. }:
\begin{align}
  &\left[(\nabla\partial_{a})-\frac{1}{d+2s-4}(a\nabla)\Box_{a}\right]\frac{I^{s}(a,b;z)}{(z^{2})^{\Delta_{(s)}}}=0\label{3.30}\\
  &\left[(\nabla\partial_{a})-\frac{1}{d+2s-4}(a\nabla)\Box_{a}\right]\frac{\tilde{t}^{(s)}(a,b,c;\hZ,\hY,\hX)}{ z^{\Delta_{(s)}} x^{\Delta_{(s)}} y^{\Delta_{(s)}}}=0\label{3.31}
\end{align}
 Another Ward identity here is tracelessness. This condition will be satisfied automatically due to the existence of traceless projection operators between correlators and fields in (\ref{3.26}) :
\begin{align}
 \Box_{a}\frac{\delta}{\delta h^{(s)}(a;x)}W^{eff}(h^{(s)})=0\quad\Leftrightarrow \quad\Box_{a}\E^{(s)}(a, b)=0\label{3.32}
\end{align}

An important point here is that all these Ward identities (conservation and tracelessness) hold only away from the singularity i.e. for $x,y,z\neq 0$.
When we  regularize our effective action (say in the dimensional regularization scheme   $d\longrightarrow d-\epsilon$), we can collect all singular terms from $\tilde{t}^{(s)}(a,b,c;\hZ,\hY,\hX)$ and then apply the general singularity-extraction formula. Investigation of the divergences these singular terms leads directly to an understanding of the trace and gauge anomalies in this higher spin case. 
This is the general strategy for investigation of conformal and gauge anomaly using correlation function and effective action. 
In the next section we present the singularity-extraction procedure.

\section{Singularity Extraction}\label{sec:four}
Here we will use and partially review some formulas  considered by  W.~R\"uhl \cite{Ruehl:2008bm} for a different purpose, connected with the local higher spin interaction. For the investigation and extraction of the singularity of the two-point function it is enough to consider the Fourier-transform formula:
\begin{align}
(x^2)^{-\lambda} &= C_{\lambda} \int dq e^{-ixq}(q^2)^{\lambda-d/2}\label{4.1}\\
C_{\lambda} &= \frac{\Gamma(-\lambda +d/2)}{2^{2\lambda} \pi^{d/2} \Gamma(\lambda)}\label{4.2}
\end{align}
It is easy to see that we can rewrite this in a form more convenient for our purposes expression
\begin{align}
(x^2)^{-\lambda} &= \tC_{\lambda} (-\Box)^{\lambda-d/2}\delta^{d}(x)\label{4.3}\\
\tC_{\lambda} &= \frac{\Gamma(-\lambda +d/2)}{2^{2\lambda-d} \pi^{-d/2} \Gamma(\lambda)}\label{4.4}
\end{align}
Then using dimensional regularization:
\begin{align}
  d\longrightarrow d-\epsilon\label{4.5}
\end{align}
 and the well-known formula for non-positive arguments of $\Gamma$ function:
\begin{align}
\Gamma(-n-\epsilon)= \frac{(-1)^{n+1}}{n!\epsilon}+ \textnormal{nonsingular part}\label{4.6}
\end{align}
we arrive at the following singularity-extraction formula for distributions:
\begin{align}
(x^2)^{-\lambda} &= \frac{\hC_{\lambda}}{\epsilon} (-\Box)^{\lambda-d/2}\delta^{d}(x)+\textnormal{nonsingular part}\label{4.7}\\
\hC_{\lambda} &= \frac{(-1)^{\lambda-d/2+1}}{2^{2\lambda-d-1}(\lambda-d/2)! \pi^{-d/2} \Gamma(\lambda)}\label{4.8}
\end{align}
In a similar but much more complicated way, we can extract the principal singularities from three point function and corresponding cubic part of effective action.
This extraction procedure one can find in \cite{Ruehl:2008bm}. We present here only the final result, reviewing the original derivation of~\cite{Ruehl:2008bm} in detail in Appendix \ref{app:B}.

The main object of interest is
\begin{align}
F_{\{\lambda\mu\nu\}}(z,x,y) = \frac{1}{(z^{2})^{\lambda}(x^{2})^{\mu}(y^{2})^{\nu}}\label{4.9}
\end{align}
The singular properties here are revealed by a Fourier transformation together with the standard methods for the evaluation of Feynman integrals.
Of course, in addition we use the same conventional method of dimensional regularization (\ref{4.5}).
%
Here we will just show final result of the extraction of the main singularity of our local distribution (see \cite{Ruehl:2008bm} and  Appendix \ref{app:B}):
\begin{align}
(2\pi)^{-d}\frac{\Omega^{\{\lambda\mu\nu\}}_{m} }{\epsilon}P^{\{\lambda\mu\nu\}}_{m}(-\Box_1,-\Box_2,-\Box_3) \delta(x_1-x_3)\delta(x_2-x_3) \label{4.10}
\end{align} 
where
\begin{align}
\Omega^{\{\lambda\mu\nu\}}_{m} =\frac{(-1)^{m+1}}{m!} 2^{-2m}(2\pi^2)^{d}\{\Gamma(\lambda)\Gamma(\mu)\Gamma(\nu)\}^{-1}\label{4.11}
\end{align}
\begin{align}
m= \lambda +\mu +\nu -d \geq 0\label{4.12}
\end{align}
and  $P^{\{\lambda\mu\nu\}}_{m}(p_1^{2},p_2^{2},p_3^{2})$ is a homogeneous  
polynomial in the $p^{2}_{i}, i=1,2,3$ of degree $m$.
We can easily evaluate this polynomial:  
\begin{align}
P^{\{\lambda\mu\nu\}}_{m}(p_1^2,p_2^2,p_3^2) = \sum_{r_1r_2r_3} R^{(m)\{\lambda\mu\nu\}}_{r_1r_2r_3}(p_1^2)^{r_1}(p_2^2)^{r_2}(p_3^2)^{r_3}\label{4.13}
\end{align}
and the rational numbers $R$ can be expressed by 
\begin{align}
&R^{(m)\{\lambda\mu\nu\}}_{r_1r_2r_3} = \delta_{m,r_1+r_2+r_3}{ m \choose r_1,r_2,r_3 } \nonumber\\ &\frac{\Gamma(r_1+r_2-\lambda+d/2)\Gamma(r_2+r_3-\mu +d/2)\Gamma(r_3+r_1-\nu+d/2)}{\Gamma(2m-\lambda-\mu-\nu+3d/2)}\label{4.14}
\end{align}
In (\ref{4.13}), the summation over $r_1, r_2, r_3$ is restricted by the requirement that the arguments of the gamma functions in (\ref{4.14}) remain positive.

\section{Singularity of Effective Action and Quantum Anomaly}\label{sec:five}
Now we are completely ready to divide our cubic effective action into singular and regular parts. First of all we should bring the three point kernel (\ref{3.23})
to the following series:
\begin{align}
  &\frac{1}{ z^{\Delta_{(s)}} x^{\Delta_{(s)}} y^{\Delta_{(s)}}}\tilde{t}^{(s)}(a,b,c;\hZ,\hY,\hX)=\sum_{\{\lambda\mu\nu\}}f_{\{\lambda\mu\nu\}}(z,y,x;a,b,c)F^{\{\lambda\mu\nu\}}(z,x,y)\label{5.1}
\end{align}
where $F^{\{\lambda\mu\nu\}}(z,x,y)$  is defined in (\ref{4.9}) and $f_{\{\lambda\mu\nu\}}(z,y,x;a,b,c)$ is a nonsingular expression, polynomial in $(az),(bz),(cz), (ay),(by),(cy),\dots , (ab),(ac),(bc)$ and the summation extends over all triplets obtained after stripping all variables 
\begin{align}
(a\hX),(b\hY),(c\hZ);I(a,b;z),I(b,c;x),I(a,c;y)\label{5.2}
\end{align}
inside of the symmetric structural tensor $\tilde{t}^{(s)}(a,b,c;\hZ,\hY,\hX)$ to the power of $z^{2},y^{2}, x^{2}$ in denominator and  nonsingular numerator $f_{\{\lambda\mu\nu\}}(z,y,x;a,b,c)$ forming different powers (triplets) of singularities $\{\lambda\mu\nu\}$. These different triplets also includes  prefactor $\frac{1}{ z^{\Delta_{(s)}} x^{\Delta_{(s)}}y^{\Delta_{(s)}}}$ as the lowest term of our series. Then we can extract singular terms using the regularization (\ref{4.5}) and rules (\ref{4.10})-(\ref{4.14}). 
\begin{align}
  &\frac{1}{ z^{\Delta_{(s)}} x^{\Delta_{(s)}} y^{\Delta_{(s)}}}\tilde{t}^{(s)}(a,b,c;\hZ,\hY,\hX)\mid _{singular}
  =\frac{1}{\epsilon}\L^{(s)}_{sing}(z,y,x;\Box_1,\Box_2,\Box_3)\delta(x_1-x_3)\delta(x_2-x_3)\nonumber\\
  &=\frac{(2\pi)^{-d}}{\epsilon}\sum_{\{\lambda\mu\nu\}}f_{\{\lambda\mu\nu\}}(z,y,x;a,b,c)\Omega^{\{\lambda\mu\nu\}}_{m} P^{\{\lambda\mu\nu\}}_{m}(-\Box_1,-\Box_2,-\Box_3) \delta(x_1-x_3)\delta(x_2-x_3) \label{5.3} 
\end{align}
Now we can split the effective action (cubic part) into singular and regular parts\footnote{To make the splitting more definite in (\ref{5.3}) and (\ref{5.4}), we moved the pole $\frac{1}{\epsilon}$ ahead of the local residue $\L^{(s)}_{sing}$}:
\begin{align}
  &W^{eff}(h^{(s)})=\frac{1}{\epsilon}W^{eff}_{sing,local}(h^{(s)})+W^{eff}_{reg,nonlocal}(h^{(s)})\nonumber\\
  &=\frac{1}{\epsilon}\int dx_{1}dx_{2}dx_{3}\L^{(s)}_{sing}\delta(x_1-x_3)\delta(x_2-x_3)*^{s}_{a}*^{s}_{b}*^{s}_{c}h^{T}(a;x_{1})h^{T}(b;x_{2})h^{T}(c;x_{3})\nonumber\\
  &\quad\quad\quad+W^{eff}_{reg,nonlocal}(h^{(s)}).\label{5.4}
\end{align}
Taking into account that our regularization is gauge invariant we can claim that our full effective action is again gauge invariant with respect to the transformation (\ref{3.29}):
\begin{align}
  &\delta W^{eff}(h^{(s)})=\frac{1}{\epsilon}\delta W^{eff}_{sing,local}(h^{(s)})+\delta W^{eff}_{reg,nonlocal}(h^{(s)})=0\label{5.5}
\end{align}
The same we can say about tracelessness:
\begin{align}
 &\Box_{a}\frac{\delta}{\delta h^{(s)}(a;x)}W^{eff}(h^{(s)})\nonumber\\
 &=\Box_{a}\frac{\delta}{\delta h^{(s)}(a;x)}\frac{1}{\epsilon}W^{eff}_{sing,local}(h^{(s)})+\Box_{a}\frac{\delta}{\delta h^{(s)}(a;x)}W^{eff}_{reg,nonlocal}(h^{(s)})=0\label{5.6}
\end{align}
On the other hand we see that due to (\ref{3.27})-(\ref{3.31}) the singular part of the transformation (\ref{5.5}) can be brought into the following equation:
\begin{align}
  &\frac{1}{\epsilon}\frac{\delta}{\delta \alpha^{(s-1)}(\ta;x_{1})} W^{eff}_{sing,local}(h^{(s)})
  =-\frac{1}{\epsilon}\E^{(s-1)}(\ta, a)*^{s-1}_{a}\left[(\nabla_{1}\partial_{a})-\frac{1}{d+2s-4}(a\nabla_{1})\Box_{a}\right]\nonumber\\
  &\times \int dx_{2}dx_{3}\L^{(s)}_{sing}(z,y,x;a,b,c;\Box_1,\Box_2,\Box_3)\delta(x_1-x_3)\delta(x_2-x_3)*^{s}_{b}*^{s}_{c}h^{T}(b;x_{2})h^{T}(c;x_{3})\label{5.7}
\end{align}
We observe that if the gauge transformation of the singular part of the effective action (or the divergence of the singular part of the correlation function) remains singular, then it must vanish according to (\ref{5.5}). This is because the transformation of the nonsingular part is finite, and no mechanism is available to cancel the singular variation. Therefore, the anomaly has a single source: the finite part of the variation of the singular local effective action. In other words, linear terms in $\epsilon$ within the variation of the singular part are required to cancel the $\epsilon$-pole in (\ref{5.7}). Thus, if
\begin{align}
\frac{1}{\epsilon}\frac{\delta}{\delta \alpha^{(s-1)}(\ta;x_{1})} W^{eff}_{sing,local}(h^{(s)})=\A(h^{(s)})\label{5.8}
\end{align}
and the right-hand side is finite, then from (\ref{5.5}) we obtain
\begin{align}
\frac{\delta}{\delta \alpha^{(s-1)}(\ta;x_{1})} W^{eff}_{reg,nonlocal}(h^{(s)})=-\A(h^{(s)})\label{5.9}
\end{align}
which provides the general characterization of a quantum anomaly. An analogous formulation applies to the trace anomaly: if the first term in (\ref{5.6}) is non-vanishing and finite, 
\begin{align}
 &\Box_{a}\frac{\delta}{\delta h^{(s)}(a;x)}\frac{1}{\epsilon}W^{eff}_{sing,local}(h^{(s)})=\T(h^{(s)})\label{5.10}
\end{align}
then it follows that
\begin{align}
 \Box_{a}\frac{\delta}{\delta h^{(s)}(a;x)}W^{eff}_{reg,nonlocal}(h^{(s)})=-\T(h^{(s)})\label{5.11}
\end{align}
Now let us explore the right-hand side of (\ref{5.7}) in detail.
When we focus our attention only on derivatives then we see that the central point of our interest here is the sum of the following expressions with
different powers of the Laplace operator
\begin{align}
  &\left[(\nabla_{1}\partial_{a})-\frac{1}{d+2s-4}(a\nabla_{1})\Box_{a}\right]\int dx_{2}dx_{3}[\Box_1]^{r_{1}}[\Box_2]^{r_{2}}[\Box_3]^{r_{3}}
  \delta(x_1-x_3)\delta(x_2-x_3)\nonumber\\
  &f(z,y,x;a,b,c)*^{s}_{b}*^{s}_{c}h^{T}(b;x_{2})h^{T}(c;x_{3})
  \label{5.12}
\end{align} 
There are two key points in this formula. First, the Laplace operators act only on delta-functions, whereas the derivative $\nabla_{1}$ also acts on the function $f(z,y,\dots)=f(x_{1}-x_{2}, x_{3}-x_{1},\dots)$. Second, there is no integration over $x_{1}$, since this corresponds to the variation of effective action on the gauge parameter $\alpha(a;x_{1})$. 
On the other hand, it is necessary to free the delta-functions from any differentiation in order to perform the integrals and obtain local expressions involving two linearized gauge fields together with a certain number of differential operators. To achieve this, one must carefully manipulate the delta-functions and partial derivatives, taking into account that, due to translation invariance, all differentiations with respect to $x_{1}$ acting on delta-functions or on the function of $z,y,x$ can be equivalently transformed into differentiations with respect to $x_{2}$ and $x_{3}$. 
For the Laplace operators the rules described above look like\footnote{For brevity we introduce $\delta_{ij}=\delta(x_{i}-x_{j}), i,j=1,2,3.$}
\begin{align}
  &\Box_{1}(\delta_{13}\delta_{23})f(z,y,x;a,b,c)\dots \nonumber\\
  &=\delta_{13}\delta_{23}\big[\Box_{1}f(z,y,\dots)+f(z,y,\dots)(\nabla_{2}
  +\nabla_{3})^{2}-2\nabla^{\mu}_{1}f(z,y,\dots)(\nabla_{2\mu}+\nabla_{3\mu})\big]\dots\label{5.13}\\
  &\Box_{2}(\delta_{13}\delta_{23})f(z,x,y;a,b,c)\dots \nonumber\\
  &=\delta_{13}\delta_{23}\big[\Box_{2}f(z,x,\dots)+f(z,x,\dots)\Box_{2}+2\nabla^{\mu}_{2}f(z,x,\dots)\nabla_{2\mu}\big]\dots\label{5.14}\\
  &\Box_{3}(\delta_{13}\delta_{23})f(y,x,z;a,b,c)\dots \nonumber\\
  &=\delta_{13}\delta_{23}\big[\Box_{3}f(y,x,\dots)+f(y,x,\dots)\Box_{3}+2\nabla^{\mu}_{3}f(y.x,\dots)\nabla_{3\mu}\big]\dots\label{5.15}
\end{align}
where we assume integration over $x_{2}, x_{3}$ and make use of translational invariance in terms of derivatives acting on functions of $z,y,x$
\begin{align}\label{5.16}
 (\nabla^{\mu}_{1}+\nabla^{\mu}_{2}+\nabla^{\mu}_{3})f(z,y,x;\dots)=0
\end{align}
The same translational invariance leads to the following action of the divergence (\ref{3.31})
\begin{align}
  &\left[(\nabla_{1}\partial_{a})-\frac{1}{d+2s-4}(a\nabla_{1})\Box_{a}\right](\delta_{13}\delta_{23})f(z,y,x;a,\dots)\dots \nonumber\\
  &=\delta_{13}\delta_{23}\Big[\partial^{\mu}_{a}f(z,y,x;a,\dots)(\nabla_{2\mu}+\nabla_{3\mu})\nonumber\\
  &-\frac{1}{d+2s-4}\Box_{a}f(z,y,x;a,\dots)[(a\nabla_{2})+(a\nabla_{3})]\Big]\dots\label{5.17}
\end{align}
We are now ready to discuss how an anomaly may arise in (\ref{5.12}). In other words, how differential operators can generate contributions linear in $\epsilon$ in order to cancel the corresponding pole in the singular terms. After some investigation we find that terms of this type arise in the divergence (\ref{5.17}) only from the trace operator
\begin{align}
\Box_{a} a^{2}=2(d-\epsilon)\label{5.18}
\end{align}
or, looking at (\ref{5.13})-(\ref{5.15}) we see other possible sources
\begin{align}
&\Box_{1}z^{2}=\Box_{1}y^{2}=-\Box_{1}(yz)=2(d-\epsilon)\nonumber\\
&\Box_{2}x^{2}=\Box_{2}z^{2}=-\Box_{2}(zx)=2(d-\epsilon)\nonumber\\
&\Box_{3}y^{2}=\Box_{3}x^{2}=-\Box_{3}(xy)=2(d-\epsilon)\label{5.19}
\end{align}
To obtain these contributions linear in $\epsilon$, we see from (\ref{5.13})-(\ref{5.15}) and (\ref{5.17}) that our translation-invariant function $f(z,y,x;a,b,c)$ must contain the corresponding terms with $a^{2}, z^{2}, y^{2}, x^{2}, (zy),(yx),(xz)$.
And the last remark we must make is the absence of a contribution to the part linear in $\epsilon$ (anomaly) of the first term divergence  operator in (\ref{5.17}) 
\begin{align}
(\nabla_{1}\partial_{a})(\delta_{13}\delta_{23})f(z,y,x;a,\dots) \dots
=\delta_{13}\delta_{23}\partial^{\mu}_{a}f(z,y,x;a,\dots)(\nabla_{2\mu}+\nabla_{\mu3})\dots\label{5.20}
\end{align}
Since the term linear in $\epsilon$ can be obtained by the operator $(\nabla_{1}\partial_{a})$, if the function $f(z,y,x;a,\dots)$ contains the terms $(az)$ or $(ay)$ similarly to (\ref{5.18}) and (\ref{5.19}), but from the right-hand side of (\ref{5.20}) or (\ref{5.17}) we see that due to translation invariance both derivatives in the spaces $a$ and $z,y$ never act on the function $f(z,y,x;a,\dots)$ simultaneously. Therefore, the source of the terms linear in $\epsilon$ in (5.12) and, accordingly, in (5.8) is the same as in (5.11) --- it is the trace of the singular part of the correlation function. This corresponds to the following picture: At the classical level we have conservation and tracelessness conditions for the HS current:
\begin{align}
  (\nabla_{x_{1}}\partial_{a})\J^{(s)}(a;x_{1})&=0 \label{5.21}\\
 \Box_{a} \J^{(s)}(a;x_{1})&=0 \label{5.22}
\end{align}
Then our consideration for the quantum effective action (\ref{5.8})-(\ref{5.12}) leads to the following anomalous behaviour of the currents at the quantum level (after singularity extraction and renormalization): 
\begin{align}
  (\nabla_{1}\partial_{a})\J^{(s)}_{R}(a;x_{1})&=\A(h^{(s)})=-\frac{1}{d+2s-4}(a\nabla_{1})\T(h^{(s)})  \label{5.23}\\
 \Box_{a} \J^{(s)}_{R}(a;x_{1})&=\T(h^{(s)}) \label{5.24}
\end{align}
Then we can shift our current:
\begin{align}
 &\tilde{\J}^{(s)}_{R}(a;x_{1})=\J^{(s)}_{R}(a;x_{1})+\frac{1}{2(d+2s-4)}a^{2}\T(h^{(s)})\label{5.25}\\
 &\Box_{a}\tilde{\J}^{(s)}_{R}(a;x_{1})=2\T(h^{(s)})\label{5.26}\\
 &(\nabla_{1}\partial_{a})\tilde{\J}^{(s)}_{R}(a;x_{1})=0\label{5.27}
\end{align}
and we see that in this way we can move the anomaly entirely into the trace, keeping the conservation condition and
thereby avoiding an HS gauge anomaly.
The crucial role in this consideration is played by the negative sign in front of the gradient term in (\ref{5.12}) and (\ref{5.17}). This produces minus in (\ref{5.23}) and restoration of the conservation condition in (\ref{5.27}). 
The final comment we should make here is the general form of the trace anomaly (\ref{5.26}) and the space-time dimension in which this anomaly should be considered.
First of all $\T(h^{(s)})$ arises from the three-point function i.e. from the cubic part of the effective action. Therefore anomaly is quadratic in linearized field and careful consideration of our constructive approach in symmetric case leads to the fact that number of derivative here is $2s$. Therefore there is a chance that it is an expression constructed from  squares of generalized HS curvature. This generalized curvatures were first introduced in \cite{deWit:1979sib} (see also \cite{Manvelyan:2010jf,Manvelyan:2007ey,Manvelyan:2007hv} for further discussion of certain aspects of the $AdS$ space background) and reproduces the usual linearized curvature for spin $s=2$. Therefore we see that the correct space-time  dimension for obtaining the trace anomaly from the three-point conformal HS correlation function is $d=4$.

\section{Conclusion and Outlook}
\indent

The constructive approach for the three-point higher spin conformal correlation function developed in our previous article is applied as a basis for the construction of the quantum effective action for corresponding higher spin conformal gauge theory. Using a symmetric version of the previously developed construction, we investigated the main singularity of the three-point function and of the effective action. We developed  general method of extraction of the main singularity and investigated singular local part of effective action. Then we considered general mechanism of production of the quantum anomaly. The benefit of this consideration is the following: this exact symmetric constructive approach to higher spin correlators appears to be the most suitable way of investigating the singularity of the correlation function, providing a route to the trace anomaly structure in the higher spin case in $d=4$. We see also that for the derivation and classification of all the emerging terms of the HS trace anomaly we need rather long and complicated computer calculations even for spin-two case. We shall address this aspect in a future publication.

\section*{Acknowledgements}

\indent

R.~M. would like to thank Stefan Theisen and Karapet Mkrtchyan for many valuable discussions during the whole period of preparation of this paper.
R.~M. and M.~K. were supported by the Science Committee of RA, within the framework of research project \# 21AG-1C060.

\section*{Appendices}
\appendix
\section{Example of spin-two}\label{app:A}
\renewcommand{\theequation}{A.\arabic{equation}}\setcounter{equation}{0}
\indent

When we look at (\ref{3.18})-(\ref{3.21}) more carefully then we see that our symmetric building blocks for the higher-spin correlation function,
obtained from the Osborn-Petkou formulation \cite{Osborn:1993cr}, \cite{Karapetyan:2023zdu}, \cite{Karapetyan:2025ick} by absorbing the proper number of inversion matrices as described in Section \ref{sec:three}, are defined in a way that is not optimal: 
\begin{align}
&\tG(a,b,c;\hZ,\hY,\hX)\nonumber\\
&=-(a\hX)I(b,c;x)-(b\hY)I(a,c;y)-(c\hZ)I(a,b;z)-2\tPsi(a,b,c;\hZ,\hY,\hX)\label{A.1}\\
&\tPsi(a,b,c;\hZ,\hY,\hX)=(a\hX)(b\hY)(c\hZ)\label{A.2}\\
&{\cal{\tF}}_{1}(a,b,c;\hZ,\hY,\hX)\nonumber\\&=[I(a,b;z)+2(a\hX)(b\hY)][I(a,c;y)+2(a\hX)(c\hZ)][I(b,c;x)+2(b\hY)(c\hZ)]\label{A.3}\\
&{\cal{\tF}}_{2}(a,b,c;\hZ,\hY,\hX)\nonumber\\
&=[I(a,b;z)+2(a\hX)(b\hY)]I(a,c;y)I(b,c;x)+I(a,b;z)[I(a,c;y)+2(a\hX)(c\hZ)]I(b,c;x)\nonumber\\
&+I(a,b;z)I(a,c;y)[I(b,c;x)+2(b\hY)(c\hZ)]\label{A.4}
\end{align}
for example in (\ref{A.1}) we have an additional contribution from (\ref{A.2}), and in (\ref{A.3}) and (\ref{A.4}) there are contributions from the ${\tG}^{2}$, ${\tG}{\tPsi}$ and ${\tPsi}^{2}$.
Therefore we can define a simpler set of building blocks, optimal for the case $s=2$.
First of all, we can get rid of the additional minus signs and introduce new symmetric spin-one terms:
\begin{align}
  &G(a,b,c;\hZ,\hY,\hX)=(a\hX)I(b,c;x)+(b\hY)I(a,c;y)+(c\hZ)I(a,b;z)\label{A.5} \\
  &\Psi(a,b,c;\hZ,\hY,\hX)=(a\hX)(b\hY)(c\hZ)\label{A.6}
\end{align}
The possible contributions to the spin-two three-point function in the symmetric formulation are:
\begin{align}
  &T_{G^{2}}=\frac{G^{2}(a,b,c;\hZ,\hY,\hX)}{(z^{2}x^{2}y^{2})^{d/2}}\label{A.7}\\
  &T_{G\Psi}=\frac{G(a,b,c;\hZ,\hY,\hX)\Psi(a,b,c;\hZ,\hY,\hX)}{(z^{2}x^{2}y^{2})^{d/2}}\label{A.8}\\
  &T_{\Psi^{2}}=\frac{\Psi^{2}(a,b,c;\hZ,\hY,\hX)}{(z^{2}x^{2}y^{2})^{d/2}}\label{A.9}\\
  &T_{I^{3}}=\frac{I(b,c;x)I(a,c;y)I(a,b;z)}{(z^{2}x^{2}y^{2})^{d/2}}\label{A.10}\\
  &T_{\hZ I+\hY I+\hX I}=\frac{[(a\hX)I(b,c;x)]^{2}+[(b\hY)I(a,c;y)]^{2}+[(c\hZ)I(a,b;z)]^{2}}{(z^{2}x^{2}y^{2})^{d/2}}\label{A.11}
\end{align}
It is obvious that all these five members have the proper symmetry. So we can now apply the conservation condition to obtain three conserved combinations from these five symmetric terms. To solve the conservation condition we should define two convenient objects:
the rescaled  divergence (for the case $s=2$) :
\begin{align}\label{A.12}
 D_{a}=(z^{2}x^{2}y^{2})^{d/2}\left[(\nabla\partial_{a})-\frac{1}{d}(a\nabla)\Box_{a}\right]
\end{align}
and, to abbreviate the notation, we set:
\begin{align}
  \hT &= (b\hY)(c\hZ)\label{A.13}\\
  \hI &=I(b,c;x)\label{A.14}
\end{align}
Then, after some complex and tedious calculations, we can write down all five divergences:
\begin{align}
 &D_a T_{G^{2}}=\left(\frac{4}{d}-2\right)\hI(a\nabla)\hT-\frac{8}{d}(a\nabla)\hT^{2}\label{A.15}\\
 &D_a T_{G\Psi}=-\frac{2}{d}\hI(a\nabla)\hT +\frac{4}{d}(a\nabla)\hT^{2}\label{A.16}\\
 &D_a T_{\Psi^{2}}=-\frac{2}{d}(a\nabla)\hT^{2}\label{A.17}\\
 &D_a T_{I^{3}}=(d - \frac{4}{d})\hI(a\nabla)\hT\label{A.18}\\
 &D_a T_{\hZ I+\hY I+\hX I}= -4 \hI(a\nabla)\hT-(d+2)(a\nabla)\hT^{2}\label{A.19}
\end{align}
After that we can easily derive our three conserved combinations:
\begin{align}
  T^{cons}_{(1)}&=T_{G^{2}} -(d-2) T_{G\Psi} - 2dT_{\Psi^{2}}\label{A.20}\\
  T^{cons}_{(2)}&=T_{I^{3}}+\frac{d^{2}-4}{2}T_{G\Psi} +(d^{2}-4)T_{\Psi^{2}}\label{A.21}\\
  T^{cons}_{(3)}&=T_{\hZ I+\hY I+\hX I} -2d T_{G\Psi} -\frac{d(d+10)}{2}T_{\Psi^{2}}\label{A.22}\\
  D_a T^{cons}_{(i)}&=0 ,\quad i=1,2,3\label{A.23}
\end{align}

\section{Detailed derivation of the singularity extraction formula}\label{app:B}
\renewcommand{\theequation}{B.\arabic{equation}}\setcounter{equation}{0}
\indent

In this appendix we give a detailed derivation of the formula for the main singularity of the three-point function presented in \cite{Ruehl:2008bm}:
We focus on the following  expression in flat space $\mathbb{R}^{d}$
\begin{align}
\mathcal{F}(x_1,x_2,x_3) = ((x_1-x_2)^{2})^{-\lambda}((x_2-x_3)^{2})^{-\mu}
((x_3-x_1)^{2})^{-\nu}\label{B.1}
\end{align}
To isolate the singularity we can use a Fourier transformation
\begin{align}
\mathcal{G}(p_1,p_2,p_3) &= \int dx_1dx_2dx_3 \mathcal{F}(x_1,x_2,x_3) \exp\left( i\sum_{i=1}^{3}x_{i}p_{i}\right) \nonumber\\
&= \delta(p_1+p_2+p_3)\Phi(p_1,p_2)\label{B.2}
\end{align}
using (\ref{4.1}) and (\ref{4.2}) we obtain 
\begin{align}
\mathcal{G}(p_1,p_2,p_3)& = C_{\lambda}C_{\mu}C_{\nu}(2\pi)^{3d} \int dq_1dq_2dq_3
(q_{1}^{2})^{\lambda -d/2}(q_{2}^{2})^{\mu-d/2}(q_{3}^{2})^{\nu-d/2}\nonumber\\
&\delta(p_1-q_1+q_3)\delta(p_2-q_2+q_1) \delta(p_3-q_3+q_2)\label{B.3}
\end{align}
Taking into account presence of translational invariants (momentum conservation):
\begin{align}
 &\delta(p_1-q_1+q_3)\delta(p_2-q_2+q_1) \delta(p_3-q_3+q_2)\nonumber\\
 &= \delta(p_1+p_2+p_3)\delta(q_2 -(p_2+q_1)) \delta(q_3-(p_{3}+p_2+q_1))\label{B.4}
\end{align}
and applying the standard formula for the evaluation of Feynman integrals::
\begin{align}
 &(q_{1}^{2})^{\lambda -d/2}(q_{2}^{2})^{\mu-d/2}(q_{3}^{2})^{\nu-d/2}\nonumber\\
 &=\frac{\Gamma(3d/2-\lambda-\mu-\nu)}{\Gamma(d/2-\lambda)\Gamma(d/2-\mu)\Gamma(d/2-\nu)}\prod^{3}_{i=1}\int ds_{i}\frac{\delta(1-\sum_{i}s_{i})s^{d/2-\lambda-1}_{1}s^{d/2-\mu-1}_{2}s^{d/2-\nu-1}_{3}}{(\sum^{3}_{i=1}s_{i}q^{2}_{i})^{3d/2-\lambda-\mu-\nu}}\label{B.5}
\end{align}
Performing the transformation from $\{q_{i}\}^{3}_{1}$ to $\{p_{i}\}^{3}_{1}$ suggested  by the delta-function restrictions (\ref{B.4}) 
\begin{align}
  \sum^{3}_{i=1}s_{i}q^{2}_{i}=Q^{2}+s_1s_2p_2^{2}+s_1s_3p_1^{2}+s_2s_3p_3^{2}\label{B.6}\\
  Q=q_{1}+s_{2}p_{2}-s_{3}p_{1},\quad\quad dq_{1}=dQ\label{B.7}
\end{align}
we can see that $dq_{i}$ integrals reduce to one $dQ$ integral
\begin{align}
&\frac{2\pi^{d/2}}{\Gamma(d/2)}\int^{\infty}_{0}\frac{Q^{d-1}dQ}{Q^{2}+A(s_{i},p_{i})^{3d/2-\lambda-\mu-\nu}},\label{B.8}\\
&A(s_{i},p_{i})=s_1s_2p_2^{2}+s_1s_3p_1^{2}+s_2s_3p_3^{2}.\label{B.9}
\end{align}
Then straightforward calculation leads to:
\begin{align}
&\int^{\infty}_{0}\frac{Q^{d-1}dQ}{Q^{2}+A^{3d/2-\lambda-\mu-\nu}}
=\frac{1}{2A^{d-\lambda-\mu-\nu}}\int^{\infty}_{0}dt\, t^{d/2-1}(t+1)^{\lambda+\mu+\nu-3d/2} \label{B.10}\\
&=\frac{1}{2A^{d-\lambda-\mu-\nu}}B(d/2,d-\lambda-\mu-\nu)=\frac{\Gamma(d/2)\Gamma(d-\lambda-\mu-\nu)}{2\Gamma(3d/2-\lambda-\mu-\nu)}
[A(s_{i},p_{i})]^{\lambda+\mu+\nu-d}\label{B.11}
\end{align}
 Now using all previous calculations we derive our Fourier transform (\ref{B.2})
\begin{eqnarray}
&&\Phi(p_1,p_2) = (2\pi^{2})^{d}2^{2(d-\lambda-\mu-\nu)}\frac{\Gamma(d-\lambda-\mu-\nu)}{\Gamma(\lambda)\Gamma(\mu)\Gamma(\nu)}\
\int_{s_{i} \geq 0}ds_1ds_2ds_3\delta(1-s_1-s_2-s_3)\nonumber\\ && \times s_1^{-\lambda+d/2-1}s_2^{-\mu +d/2-1}s_3^{-\nu+d/2-1}  \{s_1s_2p_2^{2}+s_1s_3p_1^{2}
+s_2s_3p_3^{2}\}^{\lambda+\mu+\nu-d}  \label{B.12}
\end{eqnarray}
So we see that our singularity is located in $\Gamma(d-\lambda-\mu-\nu)$.
Using the fact that $\lambda, \mu, \nu$ are natural numbers
and applying the method of dimensional regularization, i.e. replacing
\begin{equation}
d \longrightarrow d - \epsilon \label{B.13}
\end{equation}
We recognize that there exists only one series of first-order poles whenever
\begin{equation}
m= \lambda +\mu +\nu - d \geq 0 \label{B.14}
\end{equation}
Then for $\epsilon$ tending to zero and using
\begin{equation}
\Gamma(-m-\epsilon) = \Gamma(m+1+\epsilon)^{-1}\frac{\pi}{\sin\pi(m+1+\epsilon)}  
= \frac{(-1)^{m+1}}{m!} \epsilon^{-1} +O(1) \label{B.15}
\end{equation}
we obtain, as the residue of $\Phi(p_1,p_2)$ in $\epsilon$
\begin{eqnarray}
\Omega^{\{\lambda\mu\nu\}}_{m}\int ds_1ds_2ds_3 \delta(s_1+s_2+s_3-1) s_1^{-\lambda+d/2-1} s_2^{-\mu +d/2-1} s_3^{-\nu+d/2-1} \nonumber\\ \times \{s_1s_3p_1^{2} + s_1s_2p_2^{2}+s_2s_3p_3^{2}\}^{m} \mid _{p_3=-p_1-p_2} \label{B.16}
\end{eqnarray}
where
\begin{equation}
\Omega^{\{\lambda\mu\nu\}}_{m} =\frac{(-1)^{m+1}}{m!} 2^{-2m}(2\pi^2)^{d}\{\Gamma(\lambda)\Gamma(\mu)\Gamma(\nu)\}^{-1}\label{B.17}
\end{equation}

We should obtain our singularity (residue) as a local distribution in coordinate space. For that we have to take the integrals over $s_{i}, i=1,2,3$ and then perform an inverse Fourier transform, integrating with the corresponding factor:
\begin{align}
&(2\pi)^{-3d}\int dp_{1}dp_{2}dp_{3}\exp\left(-i\sum x_{i}p_{i}\right)\dots \label{B.18}
\end{align}
To do that we should first use trinomial expansion for 
\begin{align}
\{s_1s_3p_1^{2} + s_1s_2p_2^{2}+s_2s_3p_3^{2}\}^{m} \label{B.19}
\end{align}
and then use the formula for the Beta function of three arguments:
\begin{align}
 &B(\alpha_{1},\alpha_{2},\alpha_{3})=\frac{\Gamma(\alpha_{1})\Gamma(\alpha_{2})\Gamma(\alpha_{3})}{\Gamma(\alpha_{1}+\alpha_{2}+\alpha_{3})}\nonumber\\
 &=\int^{1}_{0}\int^{1}_{0}\int^{1}_{0} ds_{1} ds_{2}ds_{3}\delta(1-s_{1}-s_{2}-s_{3})s_{1}^{\alpha_{1}-1}s_{2}^{\alpha_{2}-1}s_{3}^{\alpha_{3}-1}\label{B.20}
\end{align}
Denoting the $s_{i}$ integral by $P_{m}(p_1^{2},p_2^{2},p_3^{2})$ and noting that it is a homogeneous
polynomial of degree $m$ in the three momentum squares, we find for the inverse Fourier transform
\begin{align}
(2\pi)^{-3d} \Omega^{\{\lambda\mu\nu\}}_{m}& \int dp_1dp_2dp_3 \delta(p_1+p_2+p_3) P^{\{\lambda\mu\nu\}}_{m}(p_1^2,p_2^2,
p_3^2)\exp\left(-i\sum x_{i}p_{i}\right)\nonumber\\
&= (2\pi)^{-d}\Omega^{\{\lambda\mu\nu\}}_{m}P^{\{\lambda\mu\nu\}}_{m}(-\Box_1,-\Box_2,-\Box_3) \delta(x_1-x_3)\delta(x_2-x_3) \label{B.21}
\end{align}
The homogeneous polynomial here is: 
\begin{align}
P^{\{\lambda\mu\nu\}}_{m}(p_1^2,p_2^2,p_3^2) = \sum_{r_1r_2r_3} R_{r_1r_2r_3}^{(m)\{\lambda\mu\nu\}}(p_1^2)^{r_1}(p_2^2)^{r_2}(p_3^2)^{r_3}\label{B.22}
\end{align}
and the rational numbers $R$ can be expressed by
\begin{align}
&R^{(m)\{\lambda\mu\nu\}}_{r_1r_2r_3} = \delta_{m,r_1+r_2+r_3}{ m \choose r_1,r_2,r_3 } \nonumber\\ &\times\frac{\Gamma(r_1+r_2-\lambda+d/2)\Gamma(r_2+r_3-\mu +d/2)\Gamma(r_3+r_1-\nu+d/2)}{\Gamma(2m-\lambda-\mu-\nu+3d/2)} \label{B.23}
\end{align}
To conclude this appendix, we should also note that the multi-argument Beta function is intimately connected with the general Feynman parametrization formula:
\begin{align}
\frac{1}{\prod^{n}_{i=1}A^{\alpha_{i}}_{i}}=\frac{\Gamma(\sum^{n}_{i=1}\alpha_{i})}{\prod^{n}_{i=1}\Gamma(\alpha_{i})}
\prod^{n}_{i=1}\int^{1}_{0}ds_{i}\frac{\delta(1-\sum^{n}_{i=1}s_{i})\prod^{n}_{i=1}s_{i}^{\alpha_{i}-1}}{(\sum^{n}_{i=1}s_{i}A_{i})^{\sum^{n}_{i=1}\alpha_{i}}} \label{B.24}
\end{align}
Setting $A_{i}=1, i=1,2,\dots n$, in the last expression,  we easily obtain the generalization of (\ref{B.20}):
\begin{align}
B(\alpha_{1},\dots\alpha_{n})=\frac{\prod^{n}_{i=1}\Gamma(\alpha_{i})}{\Gamma(\sum^{n}_{i=1}\alpha_{i})}=
\prod^{n}_{i=1}\int^{1}_{0}ds_{i}\delta(1-\sum^{n}_{i=1}s_{i})\prod^{n}_{i=1} s_{i}^{\alpha_{i}-1} \label{B.25}
\end{align}

\end{document}